\documentclass[11pt]{amsart}

\usepackage{a4}
\usepackage{latexsym}

\newcommand{\RR}{\mathbb{R}}
\newcommand{\NN}{\mathbb{N}}

\newcommand{\A}{\mathcal{A}}
\newcommand{\EE}{\mathbb{E}}
\newcommand{\FF}{\mathcal{F}}
\newcommand{\PP}{\mathbb{P}}

\newcommand{\supp}{{\rm supp}}
\newcommand{\inter}{{\rm int}}

\newtheorem{thm}{Theorem}[section]
\newtheorem{prp}[thm]{Proposition}
\newtheorem{lem}[thm]{Lemma}

\newtheorem{dfn}[thm]{Definition}
\theoremstyle{remark}
\newtheorem{rem}[thm]{Remark}
\newtheorem{rems}[thm]{Remarks}
\newtheorem{exmp}[thm]{Example}

\begin{document}

\title[IDS for random Schr\"odinger  
operators on manifolds]{Integrated density of states for ergodic random Schr\"odinger  
operators on manifolds}

\author{Norbert Peyerimhoff}
\author[Ivan Veseli\'c]{Ivan Veseli\'c$^1$ }
\address{Fakult\"ar f\"ur Mathematik, Ruhr-Universit\"at Bochum, Universit\"atsstr.~150, D-44780 Bochum, Germany}
\footnotetext[1]{Partially supported by the SFB 237 of the DFG and the MaPhySto-Centre of The Danish National Research Foundation.}
\email{peyerim@math.ruhr-uni-bochum.de, veselic@caltech.edu}

\date{}

\begin{abstract}
We consider the Riemannian universal covering of a compact manifold 
$M = X / \Gamma$ and assume that $\Gamma$ is amenable. We show for
an ergodic random family of Schr\"odinger operators on $X$ the
existence of a (non-random) integrated density of states. 
\end{abstract}



\maketitle

\section{Introduction and statement of results}

The integrated density of states (IDS) is an important notion in the
quantum theory of solids and describes the number of electron states
below a certain energy level per unit volume. Let us shortly explain
this notion in the case of a disordered solid, e.g., an alloy of two
metals with a crystal structure where the nuclei of the two metals are
randomly distributed at the lattice points. The situation can be
described quantum mechanically by a corresponding family $H^\omega =
\Delta + V^\omega$ of random Schr\"odinger operators. Due to the
macroscopic dimensions of the solid one can consider operators on the
whole $\RR^3$.  Let $\Lambda_n \subset \RR^3$ denote a cube of
sidelength $n$, centered at the origin, and let $H^\omega_{\Lambda_n}$
denote the restriction of $H^\omega$ to $\Lambda_n$ with a suitable
boundary condition (e.g. Dirichlet or Neumann or periodic). Then the
IDS is the limit of the corresponding eigenvalue counting functions of
$H^\omega_{\Lambda_n}$, normalized by the volume of the cubes
$\Lambda_n$. An ergodicity assumption yields the fact that one can
associate to the whole family $\{ H^\omega \}$ a non-random spectrum,
i.e., that almost all operators have the same spectrum.  Moreover, the
points of increase of the IDS coincide with the almost sure spectrum
of $\{ H^\omega \}$. The non-randomness of spectral data implies that
the alloy exhibits almost surely a particular behaviour of
conductivity. For the importance of the IDS from the viewpoint of
Solid State Physics see
\cite{Bonch-BruevichEEKMZ-1984,EfrosS-84,Lifschitz-1985,LifshitzGP-88}.
An overview over the mathematical results on random Schr\"odinger
operators is given in the books \cite{CarmonaL-90} and
\cite{PasturF-92}. Early rigorous results on the IDS can be found,
e.g., in articles by Pastur \cite{Pas-71a,Pas-71b,Pas-80} or \v Subin
\cite{Shub-79,Shub-82}. A good introductory course on random Schr\"odinger
operators is \cite{Kirsch-89a}.

Our aim is to generalize the classical existence result of a non-random
IDS for random Schr\"odinger operators to more general spaces. The main
results are Theorems \ref{main0} and \ref{main} below. They are
generalizations of \cite{PV-00}. We consider the universal
Riemannian covering $X$ of a compact Riemannian manifold $M = X /
\Gamma$ with an infinite group $\Gamma$ of deck transformations.  In
this context, Adachi, Br\"uning and Sunada \cite{BS,AdachiS-93} proved
the existence of an IDS for a $\Gamma$-periodic elliptic
operator $H$ in the case that $\Gamma$ is amenable. They
also proved that this IDS agrees with the $\Gamma$-trace of the
spectral projections of $H$. Note also that Dodziuk and Mathai in
their paper \cite{DodzM-97} on $L^2$-Betti numbers derived a result for
the IDS of the pure Laplace operator on $k$-forms. All mentioned
results on the IDS use \v Subin's \cite{Shub-82} convergence criterium based
on the Laplace-transform.

In this article we consider a family of Schr\"odinger operators
$H^\omega = \Delta + V^\omega$ (on a Riemannian manifold $X$), which are
parameterized by the elements of a probability space.  More precisely, 
we consider the following objects:

\begin{dfn} \label{setting}
  Let $X$ be the Riemannian universal covering of a compact Riemannian
  manifold $M = X /\Gamma$ and $\{ H^\omega = \Delta + V^\omega
  \}_{\omega \in \Omega}$ be a family of Schr\"odinger operators,
  parameterized by elements of the probability space $(\Omega,\A,\PP)$.
  The family $\{ H^\omega \}$ is called an {\em ergodic random family
  of Schr\"odinger operators}, if the potential $V: \Omega \times X \to
  \RR$ is jointly measurable and if there exists an {\em ergodic}
  family of measure preserving transformations $\{ T_\gamma: \Omega
  \to \Omega \}_{\gamma \in \Gamma}$ with $T_{\gamma_1 \gamma_2} =
  T_{\gamma_1} T_{\gamma_2}$ such that the potential satisfies the
  following {\em compatibility condition} 
  \begin{equation} \label{potcomp}
  V^{T_\gamma \omega}(x) = V^\omega(\gamma^{-1} x) 
  \end{equation} 
  for all $\omega \in \Omega$, $\gamma \in \Gamma$ and $x \in X$.
\end{dfn}
According to our convention $\Delta$ is a non-negative operator.

For the notion of measurability of random unbounded selfadjoint
operators we refer to \cite[Section 2]{KirschM-82a} and \cite[Chapter
V]{CarmonaL-90}. The ergodicity of such operators is thoroughly
investigated in \cite{PasturF-92}. If $X \times \Omega \ni (x,
\omega) \mapsto V^\omega(x)$ is jointly measurable, the multiplication
operator $\omega \mapsto V^\omega$ is measurable in the sense of
Kirsch-Martinelli \cite{KirschM-82a}. Furthermore, by their Proposition
2.4 we know that $H^\omega= \Delta + V^\omega$ is measurable, too.

In the Euclidean case $X = \RR^n$, already mild integrability
assumptions on $ (x, \omega) \mapsto V^\omega(x)$ ensure the
independence of the IDS of the boundary conditions (b.c.) used for its
construction (see \cite{KiMa-82}). On cubes $\Lambda_n$, one can
consider Dirichlet and Neumann b.c.~as well as periodic ones. In more
general geometries, b.c.~(in)dependence is a more subtle question. See
\cite{Sz-89,Sz-90}, where Dirichlet- and Neumann-IDS' on hyperbolic
spaces are compared.

Note that ergodicity of the family $\{ T_\gamma \}$ means that the
only invariant measurable sets $A \subset \Omega$ are measure-theoretically
trivial, i.e. $\PP(A) = 0$ or $\PP(A) = 1$. For technical reasons (in
order to apply a Sobolev lemma) we require for an ergodic
random family of Schr\"odinger operators that there exists a constant
$C_0 > $ such that
\begin{equation} \label{potreg}
  \Vert \nabla^k V^\omega \Vert_\infty \le C_0, \quad \text{for all 
$\omega \in \Omega$ and $k \le \frac{1}{2}\dim(X) +2$.}
\end{equation}
This implies in particular that $V^\omega$ is infinitesimally
$\Delta$-bounded, uniformly in $\omega$. It seems that this regularity
condition on the potential may be relaxed considerably by using
stochastic methods instead of analytical methods for the required heat
kernel estimates. Another approach to circumvent strong regularity
assumptions could be the use of quadratic forms \cite{Simon-1971}.

\begin{exmp} Let $X$ and $\Gamma$ be as in Definition \ref{setting}. Then
we can consider the following potential, which is an analogue of an
alloy-type potential in the Euclidean setting:
 
Let $u: X \to \RR$ be a smooth function with compact support. We choose
$\Omega = \times_{\gamma \in \Gamma} \RR$, equipped with the product
measure $\PP = \otimes_{\gamma \in \Gamma} \mu$, where $\mu$ is a
probability measure on $\RR$. Then the random variables $\pi_\gamma:
\Omega \to \RR$, $\pi_\gamma(\omega) = \omega_\gamma$ are independent
and identically distributed. The
transformations $(T_{\gamma_1}(\omega))_{\gamma_2} = \omega_{\gamma_1^{-1}
\gamma_2}$ are measure preserving and ergodic. Then $H^\omega = \Delta
+ V^\omega$ with
\[ 
V^\omega(x) = \sum_{\gamma \in \Gamma} \pi_\gamma (\omega) \, u(\gamma^{-1} x) 
\]
defines an ergodic random family of Schr\"odinger operators. Note that
$V^\omega$ is a superposition of $\Gamma$-translates of the single site
potential $u$ with coupling constants given by the random variables. 
\end{exmp}

Let us introduce some more notation. For a given $h > 0$, the {\em
$h$-boundary} of $D \subset X$ is defined as
\[ \partial_h D = \{ x \in X \mid d(x,\partial D) \le h \}. \]  
A subset of $X$ is called a {\em regular domain} if it is the
non-empty interior of a connected compact set with smooth boundary. A
{\em regular set} $D$ is a finite union $D = \bigcup_{j=1}^k D_j$ of
regular domains with disjoint closures $\overline{D_j}$. In the sequel
we will often deal with $h$-approximations and $h$-regularizations:

\begin{dfn} 
Let $U, V \subset X$ be open subsets and $h > 0$. $V$ is called an
{\em $h$-approximation of $U$}, if the symmetric difference satisfies
the following property:
\begin{equation} \label{reg}
U \Delta V \subset \partial_h U.
\end{equation}
If, additionally, $V$ is a regular set, we call $V$ an {\em 
$h$-regularization of $U$}. 

Similarly, a sequence $\{ V_n \}$ is called an $h$-approximation
($h$-regularization) of $\{ U_n \}$ if there is a fixed $h >
0$ with $U_n \Delta V_n \subset \partial_h U_n$ for all $n$ (and
the sets $V_n$ are regular). If only the existence and not the actual value
of $h > 0$ is of importance, we also refer to $\{ V_n \}$ as an approximation
(regularization) of $\{ U_n \}$.
\end{dfn} 

\begin{rems}

1) For $p \in X$ let $B_r(p)$ denote the open metric $r$-ball around
$p$. Then $B_{r + h}(p)$ is an $h$-approximation of $B_r(p)$, but generally not
vice versa (think, e.g., of the ball of radius $\pi-h/2$ around any point
of the unit $2$-sphere).

2) Relation (\ref{reg}) ist equivalent to $U \backslash \partial_h
U \subset V \subset (U \cup \partial_h U)$ and implies $\partial
V \subset \partial_h U$.

3) A natural procedure to construct an $h$-regularization $V$ of a set $U$ goes
as follows: Choose a smooth function $g: X \to [0,1]$ with
\[ g(x) = \begin{cases} 0, & \text{for}\ x \in X \backslash 
    U, \\
    1, & \text{for}\ x \in U \backslash \partial_h U, \end{cases} \] 
a regular value $t \in (0,1)$ of $g$, and $V = g^{-1}((t,1])$. $g$ can be
obtained by smoothing the characteristic function $\chi_{U \backslash
\partial_h U}$ via a suitable convolution process. 
\end{rems}

The restriction of a Schr\"odinger operator $H^\omega$ to a regular
set $D$ with Dirichlet boundary conditions is denoted by
$H^\omega_D$. It is well known that such an operator has discrete
spectrum and, thus, we can define the normalized eigenvalue counting
function (including multiplicities) as
\begin{equation} \label{ewcfunc}
N^\omega_D(\lambda) = \frac{\# \{ i \mid \lambda_i(H^\omega_{D}) < \lambda
 \}}{| D |}, \
\end{equation}
where $| D |$ denotes the volume of $D$. A non-negative, monotone
increasing and left-continuous function on $\RR$ is called a {\em
distribution function}. Thus, $N^\omega_D$ is a distribution function.
Note that a distribution function has at most countably many discontinuity 
points. 

For a better understanding of our general result we first state the
simpler case where we assume $\Gamma$ to be of {\em polynomial growth}. We
denote the metric open $r$-ball around $p \in X$ by $B_r(p)$. Then we
have, for every $p \in X$, a sequence of increasing radii $r_1 < r_2 <
\dots$ satisfying
\begin{equation} \label{isop0}
\lim_{n \to \infty} \frac{| \partial_d B_{r_n}(p) |}{| B_{r_n}(p) |} = 0
\quad \text{for all $d > 0$}.
\end{equation}
This follows readily from Lemma 3.2. in \cite{AdachiS-93}. Since
metric balls may not be regular sets (due to the existence of
conjugate points), we need a regularization of those balls in the
following theorem.

\begin{thm} \label{main0}
  Let $X$ be the Riemannian universal covering of a compact Riemannian
  manifold $M = X /\Gamma$ and assume that $\Gamma$ is of {\rm
  polynomial growth}. Let $H^\omega = \Delta + V^\omega$ be a family
  of ergodic random Schr\"odinger operators satisfying {\rm
  (\ref{potreg})}. Then there exists a (non-random) distribution
  function $N$ with the following property: For every $p
  \in X$ and any regularization $D_n$ of an increasing sequence of balls
  $B_{r_n}(p)$ satisfying {\rm (\ref{isop0})} we have, for almost all
  $\omega \in \Omega$, 
  \[ N(\lambda) = \lim_{n \to \infty} N_{D_n}^\omega(\lambda) \]
  at all continuity points of $N$. Note that $N_{D_n}$ denotes the
  normalized eigenvalue counting function of the restricted operator
  $H^\omega_{D_n}$ with Dirichlet boundary condition, as defined in
  {\rm (\ref{ewcfunc})}. $N$ is called the {\em integrated density of
  states (IDS)} of the family $\{ H^\omega \}$.
\end{thm}

In fact, existence of a non-random IDS can be proved in the much more
general setting of amenable covering groups $\Gamma$. In the following
general result we use the notion of ``admissible sequences''. This is
our generalization of the cubes $\Lambda_n$ in the Euclidean
case. However, to avoid too many technical details, we postpone the
precise definition of this notion to the next section.

\begin{thm} \label{main}
  Let $X$ be the Riemannian universal covering of a compact Riemannian
  manifold $M = X /\Gamma$ and assume that $\Gamma$ is {\em amenable}. Let
  $H^\omega = \Delta + V^\omega$ be a family of ergodic random
  Schr\"odinger operators satisfying {\rm (\ref{potreg})}. Then there exists
  a (non-random) distribution function $N$ such that we
  have, for every admissible sequence $D_n \subset X$ and almost every
  $\omega \in \Omega$, 
  \[ N(\lambda) = \lim_{n \to \infty} N_{D_n}^\omega(\lambda) \]
  at all continuity points of $N$. 
  $N$ is called the {\rm integrated density of states} of the family
  $\{ H^\omega \}$.
\end{thm}

As mentioned before, we do not present the definition of admissible
sequences at this point. We think it is more useful to give some
information about the existence of those sequences to give some
feeling for the applicability of Theorem \ref{main}.

\begin{prp} \label{ballprop}
  Let $X$ be the Riemannian universal covering of a compact Riemannian
  manifold $M = X /\Gamma$. For every monotone increasing sequence
  $D_n \subset X$ of regular sets satisfying the following property 
  \[
  \qquad \qquad \qquad \qquad \qquad \hfill \lim_{n \to \infty}
  \frac{| \partial_d D_n |}{| D_n |} = 0 \quad \text{for all $d > 0$},
  \qquad \qquad \qquad \qquad \text{\bf (P)} 
  \] 
  there exists a subsequence
  $D_{n_j}$ which is an admissible sequence. 
  
  Henceforth, we refer to this isoperimetric property of (not necessarily regular) subsets of
  $X$ as {\em property {\bf (P)}}. The existence of a sequence $\{ D_{n} \}$
  satisfying property {\rm {\bf (P)}} is equivalent to the fact that
  $\Gamma$ is amenable.
\end{prp}

The first part of this proposition will be proved after the definition
of admissible sequences in the next section. The equivalence-statement 
coincides essentially with \cite[Prop. 1.1.]{AdachiS-93}.

In the particular case that $\Gamma$ is  of polynomial growth
(and, thus, automatically amenable) there are two natural choices for
admissible sequences: either via {\em combinatorial balls in $\Gamma$}
or via {\em metric balls in $X$}. This is the content of Proposition
\ref{poladm} below. However, if one drops the assumption on the
polynomial growth, metric balls do not seem to be always
an appropriate choice for admissible sequences. For example, choose
$X$ to be the $3$-dimensional diagonal horosphere of the Riemannian
product of two real hyperbolic planes. $X$ is a solvable Lie group
with a left invariant metric admitting a cocompact lattice
$\Gamma$.\footnote{$X / \Gamma$ coincides with the solvmanifold
described in \cite[Example 3.8.9]{Th}.} Thus, $\Gamma$ is amenable and
Proposition \ref{ballprop} guarantees the existence of admissible
sequences.\footnote{An admissible sequence for the diagonal horosphere
is explicitely given in \cite[p. 668]{KP}.} On the other hand, metric
balls $B_r(p) \subset X$ have exponential volume growth (see
\cite[p. 669]{KP}). This yields strong evidence that a sequence of those
balls cannot be used as an admissible sequence. 

In order to state Proposition \ref{poladm} below we need, again, some
notation.

Let $X$ and $\Gamma$ be as before. It was explained in \cite[Section
3]{AdachiS-93} how to obtain a connected polyhedral
{\em $\Gamma$-fundamental domain} $\FF \subset X$ by lifting simplices of a
triangularization of $M$ in a suitable manner. $\FF$ consists of
finitely many smooth images of simplices which can overlap only at
their boundaries.  Using a polyhedral fundamental domain $\FF$, any
finite subset $I \subset \Gamma$ induces naturally a corresponding
subset $\phi(I) \subset X$ defined as
\begin{equation} \label{fi}
\phi(I) = \inter (\overline{I \FF}) = 
\inter(\bigcup_{\gamma \in I} \overline{\gamma \FF}).  
\end{equation}

\begin{prp} \label{poladm}
  Let $X$ be the Riemannian universal covering of a compact Riemannian
  manifold $M = X /\Gamma$ and $\Gamma$ be of polynomial growth. 

  a)Let   $\FF$ be a connected polyhedral fundamental domain and $\phi$ be the
  corresponding map (see {\rm (\ref{fi})}). Let $e$ be the identity element of $\Gamma$, $E$ a finite set of
  generators of $\Gamma$ with $e \in E = E^{-1}$ and  $E^n \subset \Gamma$ the combinatorial ball of
  radius $n \in \NN$ about $e$. Then there exists an increasing sequence $r_1 < r_2 < \dots$ of integer radii such that
  \begin{equation} 
  \label{ballquot1} 
  \frac{\vert E^{r_n+d} \backslash  E^{r_n-d} \vert}{\vert E^{r_n} \vert} \to 0 \quad \text{for all $d
  \in \NN$.}  
  \end{equation}
  Any regularization of $\{ \phi(E^{r_n}) \}$ is an admissible
  sequence.

  b) Let $p \in X$ be an arbitrary point. Then there exists an increasing
  sequence $r_1 < r_2 < \dots$ of radii such that the corresponding 
  metric balls $\{ B_{r_n}(p) \}$ satisfy {\rm (\ref{isop0})}. Moreover, 
  any regularization of $\{ B_{r_n} \}$ is an admissible sequence.
\end{prp}

We obtain as an immediate consequence of Proposition \ref{poladm} b) that 
Theorem \ref{main0} is a particular case of Theorem \ref{main}. Thus it 
suffices to prove Theorem \ref{main} which is done in Section 4. 

{\bf Acknowledgements:} We would like to thank Uwe Abresch, Jozef
Dodziuk, Werner Kirsch, Gerhard Knieper, Daniel Lenz, Thomas Schick
and Anton Thalmaier for many discussions concerning this paper. In
particular, we are grateful to J. Dodziuk for helpful remarks about
heat kernel estimates and to Th.~Schick for the useful reference to
Linderstrauss' paper \cite{Lindenstrauss-99}.

\section{Admissible sequences and ergodic theorem}

An important tool in the existence proof of a {\em non-random} IDS is an
ergodic theorem for the group $\Gamma$ of deck transformations on
$X$. We will use Lindenstrauss' pointwise ergodic theorem which is
related to a maximal ergodic theorem of Shulman (see \cite{Shul-88} and
\cite{Lindenstrauss-99}; further informations about ergodic theorems
can be found in \cite{Krengel-85} or \cite{Tempelman-92}). Lindenstrauss' 
theorem applies to discrete amenable groups. This section contains some
useful geometric facts and their interaction with this ergodic theorem. 

As in the previous section, let $\FF \subset X$ denote a connected polyhedral 
fundamental domain of $\Gamma$ and $\phi$ the associated map from finite
subsets of $\Gamma$ to open subsets of $X$ (see (\ref{fi})), which we assume 
to be fixed once and for all.

\begin{dfn} \label{admseq}
A sequence $\{ D_n \}$ of regular subsets of $X$ is called an 
{\em admissible sequence of $X$} if the following properties are satisfied:
\begin{itemize}
  \item There exists a sequence $\{ I_n \}$ of monotone increasing,
  non-empty, finite subsets of $\Gamma$ with \begin{eqnarray} \lim_{n
  \to \infty} \frac{\vert I_n \Delta I_n \gamma \vert} {\vert I_n
  \vert} &=& 0, \quad \text{for all $\gamma \in \Gamma$},
  \label{folner} \\ \sup_{n \in \NN} \frac{\vert I_{n+1} I_n^{-1}
  \vert}{\vert I_{n+1} \vert} &<& \infty. \label{doubling}
  \end{eqnarray} 
  Let $A_n = \phi(I_n)$. (Lemma \ref{equiv} below implies that $\{ A_n \}$ 
  satisfies property {\rm {\bf (P)}}.)
  \item Either $\{ A_n \}$ is an approximation of $\{ D_n \}$ and 
  $\{ D_n \}$ satisfies the isoperimetric property {\rm {\bf (P)}}, or
  $\{ D_n \}$ is an approximation of $\{ A_n \}$. (In the second case,
  $\{ D_n \}$ satisfies property {\rm {\bf (P)}} automatically, see the second
  statement of Lemma \ref{isopher} below.)
\end{itemize}
  Sequences satisfying only {\rm (\ref{folner})} are called {\em
  F{\o}lner sequences}. Monotone increasing sequences $\{ I_n \}$
  satisfying {\rm (\ref{folner})} and {\rm (\ref{doubling})} are
  called {\rm tempered F{\o}lner sequences}.
\end{dfn}

(\ref{folner}) describes geometrically that the group $\Gamma$ is
amenable.  Condition (\ref{doubling}) and the notion `tempered
F{\o}lner sequence'' are due to A. Shulman who proved a maximal
ergodic theorem for those sequences. The following proposition states
that (\ref{doubling}) is not a serious restriction for F{\o}lner
sequences:

\begin{prp}[see {\cite[Prop. 1.5]{Lindenstrauss-99}}] \label{Liprop}
  Every F{\o}lner sequence has a tempered subsequence. In particular,
  every amenable group admits a monotone increasing sequence $\{ I_n \}$
  satisfying {\rm (\ref{folner})} and {\rm (\ref{doubling})}.
\end{prp}

Next we state Lindenstrauss' pointwise ergodic theorem 
{\cite[Thm. 1.3]{Lindenstrauss-99}}.

\begin{thm} \label{liergthm}
  Let $\Gamma$ be an amenable discrete group and $(\Omega,\A,\PP)$ be
  a probability space. Assume that $\Gamma$ acts {\em ergodically} on
  $\Omega$ by measure preserving transformations $\{ T_\gamma \}$. Let
  $\{ I_n \}$ be a tempered F{\o}lner sequence. Then we have, for
  every $f \in L^1(\Omega)$ and for almost all $\omega \in \Omega$,
  \begin{equation} \label{aver} \lim_{n \to \infty} \frac{1}{\vert I_n
  \vert} \sum_{\gamma \in I_n^{-1}} f(T_\gamma \omega) = \EE(f) =
  \int_\Omega f(\omega) d\PP(\omega).  \end{equation} Furthermore 
  \[ \lim_{n \to \infty} \int_\Omega \Big | \EE(f) -\frac{1}{\vert I_n
  \vert} \sum_{\gamma \in I_n^{-1}} f(T_\gamma \omega) \Big |
  d\PP(\omega) = 0 . \] 
%  The convergence is also true in the $L^1$-norm (see \cite{Krengel-85}).
\end{thm}
In the statement of the theorem one can replace the space $L^1$ by
$L^2$, due to Shulman \cite{Shul-88}. Mean ergodic theorems hold in
more general circumstances, see, e.g., \cite[Thm. 6.4]{Tempelman-72} or
\cite[\S 6.4]{Krengel-85}.

The reader might wonder why there is a summation over $I_n^{-1}$
instead of $I_n$ in (\ref{aver}). The reason for this choice is simply
that we want it to fit, without modification, for the application
later in the paper. Lindenstrauss' theorem contains a summation over
$I_n$. Accordingly, our conditions on $I_n$ agree with those of him
only after replacing $I_n$ by $I_n^{-1}$. Note that condition
(\ref{folner}) is equivalent to
\[ 
\lim_{n \to \infty} \frac{\vert I_n^{-1} \Delta \gamma I_n^{-1}\vert} 
{\vert I_n^{-1} \vert} = 0, \quad \text{for all $\gamma \in \Gamma$}.
\]

The following lemma exhibits a useful connection between the isoperimetric 
property {\bf (P)} and the F{\o}lner condition (\ref{folner}). 

\begin{lem} \label{equiv}
  Let $I_n \subset \Gamma$ be a sequence of non-empty, finite sets
  and let $A_n = \phi(I_n)$. Then the following properties are equivalent:
  \begin{itemize}
  \item[a)] $\{ I_n \}$ satisfies the F{\o}lner condition {\rm (\ref{folner})}.
  \item[b)] $\{ A_n \}$ satisfies the isoperimetric property {\rm {\bf (P)}}.
  \end{itemize}
\end{lem}

\begin{proof} We first show that a) implies b). For an arbitrary fixed $d > 0$ we 
define the following finite set:
\[ B = \{ g \in \Gamma \mid d(\overline{g \FF},\overline{\FF}) \le d \}. \]
We first observe that if $A = \overline{\phi(I)}$ then
\[ T_d(A) := \{ x \in X \mid d(x,A) \le d \} \subset \overline{\phi(IB)}. \]
In fact, for any $x \in T_d(A)$ there exists an $x_0 \in A$ and a
$\gamma \in I$ with $d(x,x_0) \le d$ and $x_0 \in \overline{\gamma
\FF}$. Consequently, we have $d(\gamma^{-1}x,\overline{\FF}) \le d$ and, thus,
there exists a $g \in B$ with $\gamma^{-1}x \in \overline{g\FF}$. This
implies $x \in \overline{\gamma g \FF} \subset \overline{\phi(IB)}$.

For the proof we apply this observation twice. From $A_n = \phi(I_n)$
we conclude that
\[ T_d(\overline{A_n})\, \backslash\, A_n\, \subset\, 
\overline{\phi(I_n B)}\, \backslash\, \phi(I_n)\, =\, 
\overline{ \phi(I_n B \backslash I_n) }.
\]
Let $H_n = I_n B \backslash I_n$. A second application of the above observation
yields
\begin{eqnarray*}
\partial_d A_n 
&\subset& T_d( T_d(\overline{A_n})\, \backslash\, A_n )\ 
\subset \ \overline{\phi(H_n B)} 
\\
& &
\\
&\subset& \overline{\phi( \bigcup_{g_1, g_2 \in B} (I_n g_1 \Delta I_n) g_2) }\ 
=\ \bigcup_{g_1, g_2 \in B} \overline{\phi((I_n g_1 \Delta I_n)g_2) }.
\end{eqnarray*}
This implies
\[ \frac{| \partial_d A_n |}{| A_n |} \le \sum_{g_1, g_2 \in B} 
\frac{| \phi(I_n g_1 g_2 \Delta I_n g_2) |}{| \phi(I_n) |} = | B | \cdot
\sum_{g_1 \in B} \frac{| I_n g_1 \Delta I_n |}{| I_n |} \longrightarrow 0, \]
finishing the proof of the first implication.

For the proof of ``b) $\Rightarrow$ a)'' it suffices to show that there is a
$d > 0$ (not dependent on $n$) such that $\phi(I_n \Delta I_n \gamma)
\subset \partial_d A_n$. We first prove that $\phi(I_n \gamma
\backslash I_n) \subset \partial_{d_0 + d_1} A_n$, where $d_0 =
d(\overline{\gamma \FF}, \overline{\FF})$ and $d_1 = {\rm
diam}(\overline{\FF})$. Let $g \in I_n \gamma \backslash I_n$. Then $g
= g_0 \gamma$ with $g_0 \in I_n$ and $g_0 \FF \subset A_n$, $g \FF
\cap A_n = \emptyset$. Since $d(\overline{g \FF},\overline{g_0 \FF}) = d_0$
we conclude that there exists a $z \in \partial A_n$ with $d(z,\overline{g 
\FF}) \le d_0$. This implies that 
\[ g \FF \subset \partial_{d_0+d_1} A_n, \]
finishing this inclusion. We are done if we prove that $\phi(I_n
\backslash I_n \gamma) \subset \partial_{2 d_0 + 2 d_1} A_n$. Let $g \in
I_n \backslash I_n \gamma$. Then we have $g \gamma^{-1} \in I_n
\gamma^{-1} \backslash I_n$ and we obtain by the previous
considerations that
\[ g \gamma^{-1} \FF \subset \partial_{d_0+d_1} A_n. \]
The required inclusion follows now from $g \FF \subset T_{d_0+d_1}(g
\gamma^{-1} \FF)$. \end{proof}

\begin{lem} \label{isopher}
  Let $\{ U_n \}$ be a sequence of subsets of $X$ satisfying the
  isoperimetric property {\rm {\bf (P)}}. Then we have, for every radius $r
  > 0$, an index $n_0 = n_0(r)$ such that every set $U_n$, $n \ge n_0$,
  contains a metric ball of radius $r$. 

  Moreover, if $\{ V_n \}$ is an approximation of $\{ U_n \}$, then
  $\{ V_n \}$ satisfies also property {\rm {\bf (P)}}.
\end{lem}

\begin{proof} We assume that there exists an $r > 0$ and a sequence $n_1 <
n_2 < n_3 < \dots$ such that we have $B_r(p) \not\subset U_{n_j}$ for
all $p \in U_{n_j}$ and all $j \ge 1$. This implies $U_{n_j} \subset
\partial_r U_{n_j}$, which is a contradiction to property {\bf (P)}. It remains
to prove the second statement. We have $U_n \backslash \partial_h U_n
\subset V_n \subset (U_n \cup \partial_h U_n)$ and $\partial V_n \subset
\partial_h U_n$. This implies that
\begin{equation} \label{isopest}
\frac{| \partial_d V_n |}{|V_n|} \le \frac{| \partial_{h+d} U_n |}{| U_n |}
\frac{| U_n |}{|U_n \backslash \partial_h U_n|} = \frac{|
\partial_{h+d} U_n |}{| U_n |} \frac{| U_n |}{| U_n | - {| U_n \cap
\partial_h U_n |}} \to 0.
\end{equation}
Note that, for $n$ large enough, the denominator $|U_n \backslash \partial_h 
U_n|$ in (\ref{isopest}) is strictly larger than $0$. \end{proof}

\begin{rem} The second statement in Lemma \ref{isopher} is {\bf not} symmetric w.r.t.
$V_n$ and $U_n$.
\end{rem}   

\medskip

Proposition \ref{ballprop} is now a consequence of Proposition \ref{Liprop}
and the previous geometric considerations:

\begin{proof}[Proof of Proposition \ref{ballprop}]
Let $D_n \subset X$ be as in the proposition. We define 
\[ I_n = \{ \gamma \in \Gamma \mid \gamma \FF \subset D_n \}. \]
Note that $I_n \subset \Gamma$ is monotone increasing and non-empty,
for $n$ sufficiently large, by the first statement of Lemma
\ref{isopher}. One easily checks that $A_n = \phi(I_n)$ is a
$d_1$-approximation of $D_n$ with $d_1 = {\rm
diam}(\overline{\FF})$. Thus, by the second statement of Lemma
\ref{isopher}, $A_n$ inherits the isoperimetric property of
$D_n$. This implies, by Lemma \ref{equiv}, that $I_n$ is an increasing
F{\o}lner sequence. By Proposition \ref{Liprop}, there exists a
tempered F{\o}lner subsequence $I_{n_j}$. Consequently, $D_{n_j}$ is
an admissible sequence. \end{proof}

\begin{proof}[Proof of Proposition \ref{poladm}] Note that
$\Gamma$ is of polynomial growth. We first prove a). The existence of
an increasing sequence of radii $r_n$ satisfying (\ref{ballquot1}) was
proved in \cite[Prop. 5]{Adachi-93}. Let $I_n = E^{r_n}$. (\ref{folner}) 
follows readily from Lemma \ref{equiv}. By Gromov's famous result \cite{Gr},
$\Gamma$ is almost nilpotent and this implies, together with \cite{Ba},
that there exist a constant $C \ge 1$ such that
\begin{equation} \label{polycomp}
\frac{r^k}{C} \le | E^r | \le C r^k \quad \text{ for all } r \in \NN
\end{equation}
where $k \in \NN$ is the degree of $\Gamma$. This immediately yields
(\ref{doubling}):
\[ \frac{| I_{n+1} I_n^{-1} |}{| I_{n+1} |} \le \frac{| E^{2 r_{n+1}} |}
{| E^{r_{n+1}} |} \le 2^k C^2. \]

Now we prove b). W.l.o.g. we can assume that $p \in
\overline{\FF}$. Let $I_n = \{ \gamma \in \Gamma \mid \gamma \FF
\subset B_{r_n}(p)$. The F{\o}lner property of $I_n$ follows precisely
as in the proof of Proposition \ref{ballprop}. It remains to prove
(\ref{doubling}). One easily checks that
\begin{equation} \label{BAB}
B_{r_n-d_1}(p) \subset A_n = \phi (I_n) \subset B_{r_n}(p)
\end{equation}
with $d_1 = {\rm diam}(\overline{\FF})$. Let $\Vert \cdot \Vert$ denote the 
word norm of $\Gamma$ with respect to $E$. Milnor showed in \cite{Milnor} 
that there are $a \ge 1$, $b \ge 0$ such that
\[ \frac{1}{a}\Vert \gamma \Vert - b \le d(p,\gamma p) \le a \Vert \gamma 
\Vert.\]
This implies, together with (\ref{BAB}), that
\[ \{ \gamma \in \Gamma \mid \Vert \gamma \Vert \le \frac{1}{a}(r_n-d_1) \} 
\subset I_n \subset \{ \gamma \in \Gamma \mid \Vert \gamma \Vert \le a
(r_n+b) \}, \]
and the same inclusions hold for $I_n^{-1}$. Consequently, we have
\[ \{ \gamma \in \Gamma \mid \Vert \gamma \Vert \le \frac{2}{a}(r_n-d_1) \} 
\subset I_n^{\,} I_n^{-1} \subset \{ \gamma \in \Gamma \mid \Vert \gamma \Vert \le 
2 a (r_n+b) \}, \]
and the required estimate follows, again, by (\ref{polycomp}). \end{proof}

The final lemma of this section, which we will apply later to heat kernels, 
is an immediate consequence of Lindenstrauss' ergodic theorem.

\begin{lem} \label{Lindenstrausslem}
  Let $I_n \subset \Gamma$ be a tempered F{\o}lner sequence and $A_n =
  \phi(I_n) \subset X$. Assume that $\Gamma$ acts ergodically on
  a probability space $(\Omega,\A,\PP)$ by measure preserving
  transformations $\{ T_\gamma \}$. Let $f : \Omega \times X \to \RR$
  be a jointly measurable bounded  function satisfying the
  {\em compatibility condition}
  \[ f(T_\gamma \omega,x) = f(\omega, \gamma^{-1} x) \]
  for all $\omega \in \Omega$, $\gamma \in \Gamma$
  and $x \in X$. Then we have, for almost all $\omega \in \Omega$,
  \begin{equation} \label{Lindenstrausslim} 
  \lim_{n \to \infty} \frac{1}{| A_n |} \int_{A_n} f(\omega, x)\, dx = 
  \frac{1}{| \FF |} \EE \left( \int_\FF f ( \bullet , x)\, dx \right),
  \end{equation}
  where $\EE$ denotes the expectation on $\Omega$. The convergence holds
  in the $L^1(\Omega)$-topology, as well.
\end{lem}

\begin{proof} Let $F(\omega) = \int_\FF f(\omega,x) dx$. Then we obviously 
have $F \in L^1(\Omega)$. We conclude that
\begin{eqnarray*}
\frac{1}{| A_n |} \int_{A_n} f(\omega,x)\, dx &=& \frac{1}{
\vert I_n \vert}  \sum_{\gamma \in I_n} \frac{1}{\vert \FF \vert}
\int_{\gamma \FF} f(\omega,x)\, dx \\
&=& \frac{1}{ \vert I_n \vert} 
\sum_{\gamma \in I_n}  \frac{1}{\vert \FF \vert}
\int_{\FF} f(\omega,\gamma x)\, dx \\
&=& \frac{1}{ \vert I_n \vert}  
\sum_{\gamma \in I_n} \frac{1}{\vert \FF \vert}
\int_{\FF} f(T_{\gamma^{-1}} \omega,x)\, dx \\
&=& \frac{1}{ \vert I_n \vert} 
\sum_{\gamma \in I_n^{-1}} \frac{1}{\vert \FF \vert}
F(T_\gamma \omega).
\end{eqnarray*} 
Now Theorem \ref{liergthm} implies
\[ \lim_{n \to \infty} \frac{1}{| A_n |} \int_{A_n} f(\omega,x) dx = \frac{1}{| \FF |} 
\EE(F), \]
for almost all $\omega \in \Omega$ and in $L^1$-sense. \end{proof}

\section{Heat kernel estimates}

In this section we derive heat kernel estimates for a family of
Schr\"odinger operators $\{ H^\omega = \Delta + V^\omega \}_{\omega \in \Omega}$ satisfying the regularity condition
\begin{equation} \label{potreg1}
  \Vert \nabla^k V^\omega \Vert_\infty \le C_0, \quad \text{for all 
$\omega \in \Omega$ and $k \le \frac{1}{2}\dim(X) +2$.}
\end{equation}
These estimates are, besides an ergodic theorem, the second crucial
tool for our existence proof of an IDS.

Due to the Kato-Rellich Theorem \cite{ReedS-II} all $H^\omega$ are  densely defined selfadjoint operators on
$L^2(X)$ and their domains coincide.
By the spectral theorem we can define the operator $\exp (-t
H^\omega)$, which has an integral kernel $k^\omega (t,\cdot,
\cdot)$. Let $D \subset X$ be a regular set. We denote
the restriction of $H^\omega$ to $D$ with Dirichlet boundary
conditions by $H^\omega_D$ and the corresponding heat kernel of
$\exp(-t H_D^\omega)$ by $k_D^\omega (t,\cdot, \cdot)$. We will need
the following estimates.

\begin{prp} 
  \label{kernellem} Let $H^\omega = \Delta + V^\omega$, $\omega \in
  \Omega$, be a family of operators satisfying {\rm (\ref{potreg1})}. Then
  the following estimates hold:
  \begin{itemize} 
  \item[a)] {\bf Domain Monotonicity:} 
  For every regular set $D \subset X$ we have \[ 0 \le
  k^\omega_D(t,x,y) \le k^\omega(t,x,y), \] for all $x,y \in D$ and $t
  > 0$.
  \item[b)] {\bf Upper Bound:} 
  There exists a function $C(t)$, $t > 0$, such that \[ 0
  \le k^\omega(t,x,y) \le C(t), \] for all $x,y \in X$ and $\omega \in
  \Omega$.
  \item[c)] {\bf Principle of not feeling the boundary:} 
  For all $t > 0$ there exists an $h =h(t,\epsilon) > 0$ such that, for all
  regular sets $D \subset X$ and all $\omega \in \Omega$, we have \[
  \vert k^\omega(t,x,y) - k^\omega_D(t,x,y) \vert \le \epsilon \quad
  \text{for all $x,y \in D \backslash \partial_h D$}. \] \end{itemize}
\end{prp}

\begin{proof} 
  Inequality a) is a consequence of the maximum principle for solutions of the
  heat equation (see, e.g., \cite{Taylor-1996} or \cite{Cha}). 

  We consider now assertion b). Let $t > 0$ be fixed. The heat kernel $k(t,x,y)$ of the Laplacian on
  $X$ (i.e., without potential) is a continuous function (see, e.g., \cite{Dav-90}) satisfying
  $k(t,x,y) = k(t,\gamma x,\gamma y)$ for any $\gamma \in \Gamma$.
  Since $\Gamma$ acts cocompactly on $X$ we conclude the existence of
  a constant $C_1(t)$ with $0 \le k(t,x,x) \le C_1(t)$. A simple
  application of the semigroup property yields the same off-diagonal
  estimate 
  \[ 0 \le k(t,x,y) \le \sqrt{k(t,x,x) k(t,y,y)} \le C_1(t). \] 
  The potential can be treated with stochastic arguments which was proposed
  to us by A. Thalmaier: Since $X$ is stochastically complete,
  we can apply the Feynman-Kac formula for manifolds (see, e.g.,
  \cite{Elworthy}) and obtain, for every $f \in C_0^\infty(X)$: 
  \[
  \int_X k^\omega(t,x,y) f(y) dy = \EE_x \Big(\, f(b_t)\exp(\int_0^t
  V^\omega(b_s) ds) \, \Big), 
  \] 
  where $b_t$ is the Brownian motion on $X$
  starting in $x$ and $\EE_x$ is the corresponding expectation. Using
  $\Vert V^\omega \Vert \le C_0$ we obtain for every non-negative $f
  \in C_0^\infty(X)$: \[ \vert \int_X k^\omega(t,x,y) f(y) dy \vert \,
  \le \, \EE_x(\, f(b_t) \, ) e^{C_0 t} \, \le \, C_1(t) e^{C_0 t}
  \Vert f \Vert_1.  \] Continuity of $k^\omega$ implies \[ \vert
  k^\omega(t,x,y) \vert \le C_1(t) e^{C_0 t}, \] finishing the proof
  of b).
  
  The proof of c) is based on finite propagation speed of the wave equation.
  The roots of this approach can be found in \cite{CHGrT-1982}. We follow the
  arguments given in \cite[Thm 2.26]{LueckS-1999} and which are attributed to
  U.~Bunke. For the reader's convenience we present the proof in detail (see
  also \cite{DodzM-97} for a related method).

  To simplify the notation we omit the index $\omega$. Due to condition 
  (\ref{potreg1}), all inequalities hold uniformly in $\omega \in \Omega$. 
  
  In what follows, $t > 0$ is fixed and $h > 0$ is kept variable. Let $x_0,
  y_0 \in D^h := D \backslash \partial_h D$ and $B_1 = B_{h/3}(x_0)$ and $B_2 = B_{h/3}(y_0)$ be the
  corresponding balls. Our first aim is to prove existence of a function
  $C(h)$ with $C(h) \to 0$, as $h \to \infty$ such that, for every $u \in
  C_0^\infty(B_2)$ and $f = (e^{-t H} - e^{-t H_D})u$, the following pointwise
  estimate holds:
  \begin{equation} \label{pointest}
  \vert H^k f(x_0) \vert \le C(h) \Vert u \Vert_2.
  \end{equation}
  Our departure point is the following Fourier transform identity
  \begin{equation} \label{fourier} 
  \frac{(-1)^m}{\sqrt{\pi t}}\int_0^\infty \left( \frac{d^{2m}}{ds^{2m}}
    e^{-s^2/4t} \right) \cos(s\xi) ds = \xi^{2m}e^{-t\xi^2}. 
  \end{equation}  
  Applying the spectral theorem to (\ref{fourier}) with $\xi = \sqrt{H}$ and
  $\xi = \sqrt{H_D}$ we obtain
  \[ H^{k+l}f = \int_0^\infty P(s) e^{-s^2/4t}\, 
  (\cos(s\sqrt{H})-\cos(s\sqrt{H_D}))u \,ds,
  \]     
  where $P(s)$ is a fixed polynomial. Note that the coefficients of $P$ are
  expressions in $t$ and that $t$ is considered as a fixed positive constant.
  Unit propagation speed (see, e.g., \cite{Taylor-1996}) implies, for $g_s =
  \cos(s\sqrt{H})u$ and $h_s=\cos(s\sqrt{H_d})u$ that
  \[ \supp(g_s),\, \supp(h_s) \subset B_{2h/3}(y_0) \subset D \]
  for $s < h/3$. Since $g_s$ and $h_s$ both satisfy the wave equation with
  initial conditions $g(0,\cdot) =u$, $\frac{\partial g}{\partial s}(0,\cdot)
  = 0$, we conclude that $g_s - h_s \equiv 0$, for $0 < s < h/3$. 
  The Cauchy-Schwarz inequality yields
  \begin{eqnarray*}
  \Vert H^{k+l}f \Vert_{L^2(B_1)}^2 &\le& \int_{B_1} \left( \int_{h/3}^\infty
  \vert P(s) (g_s(x) - h_s(x)) \vert\, e^{-s^2/4t} ds \right)^2 dx \\
  &\le& A_1(h) \int_{h/3}^\infty \vert P(s) \vert e^{-s^2/4t}
  \int_{B_1} (g_s(x) - h_s(x))^2 dx\, ds,
  \end{eqnarray*}
  where $A_1(h) = \int_{h/3}^\infty \vert P(s)\vert e^{-s^2/4t} ds \to 0$, as
  $h \to \infty$. Using, again, the spectral theorem, we conclude from $\vert
  \cos(s\xi)\vert \le 1$ that
  \[ \Vert H^{k+l}f \Vert_{L^2(B_1)} \le 2 A_1(h) \Vert u \Vert_{L^2(B_2)}. \]
  
  In order to obtain the pointwise estimate (\ref{pointest}), we would like
  to apply a Sobolev inequality of the type
  \begin{equation} \label{sobolev}
  \vert g(x) \vert \le \sum_{l=0}^N a_l \Vert H^l g \Vert_{L^2(B_{h/3}(x))}
  \end{equation} 
  for all $g \in C_0^\infty(B_{h/3}(x))$, where $N = [ \frac{\dim X}{2} +2 ]$,
  and the coefficients $a_l$ are independent of $x \in X$ and $\omega \in
  \Omega$. This is possible since $X/\Gamma$ is compact. Moreover, the
  condition $g \in C_0^\infty(B_{h/3}(x))$ can be relaxed to $g \in
  C^\infty(B_{h/3}(x))$, since, for $h \ge h_0 > 0$, we can choose, for every
  point $x \in X$, cut-off functions $\rho_x \in C_0^\infty(B_{h/3}(x))$ with
  universal bounds on the derivatives and, thus, apply an estimate
  \[ 
  \Vert H^l \rho_x g \Vert_{L^2(B_{h/3}(x))} \le \beta_l\, \Vert g 
  \Vert_{L^2(B_{h/3}(x))} + \gamma_l\, \Vert H^l g \Vert_{L^2(B_{h/3}(x))} 
  \] 
  with universal constants $\beta_l, \gamma_l$. This proves (\ref{pointest}), 
  namely
  \[ \vert H^k f(x_0) \vert \le \sum_{l=0}^N a_l\, \Vert H^{k+l} f
  \Vert_{L^2(B_1)} \le A_2(h)\, \Vert u \Vert_{L^2(B_2)}, \]
  where $A_2(h) \to 0$, as $h \to \infty$. 

  Next the heat kernels come into play:
  \begin{gather*}
    \vert \langle H_y^k(k(t,x_0,\cdot)-k_D(t,x_0,\cdot)), u \rangle_{L^2(B_2)}
    \vert
    \hspace{6cm} \\
    = \vert \int_{B_2} H_y^k(k(t,x_0,y) - k_D(t,x_0,y))\, u(y)\, dy \vert \\
    = \vert \int_{B_2} (k(t,x_0,y) - k_D(t,x_0,y))\, (H^k u)(y)\, dy \vert \\
    = \vert ((e^{-tH}-e^{-tH_D})H^ku)(x_0) \vert \\
    \hspace{7cm} = \vert H^k f(x_0) \vert \le A_2(h) \Vert u \Vert_{L^2(B_2)}.
  \end{gather*}
  Since $u \in C_0^\infty(B_2)$ was arbitrary, we conclude that 
  \[ \Vert H_y^k(k(t,x_0,\cdot)-k_D(t,x_0,\cdot)) \Vert_{L^2(B_2)} \le
  A_2(h). \] 
  Again, using the Sobolev inequality (\ref{sobolev}), we end up with
  \[ \vert k(t,x_0,y_0) - k_D(t,x_0,y_0) \vert \le A_3(h), \]
  where $A_3(h) \to 0$, as $h \to \infty$. Choosing $h$ large enough, we
  obtain the required estimate of the lemma. \end{proof}
  
\begin{rems} 
  
1) Estimate b) of Proposition \ref{kernellem} is very crude, but sufficient for our
purposes. For a much better estimate, we refer the reader to \cite{LiY-86}.
 
2) For the Neumann heat kernel, estimate c) is still valid \cite{LueckS-1999}. However, Domain Monotonicity for Neumann heat kernels is a subtle question. See \cite{Chavel-1986,BAssB-1993} and \cite{CarmonaZ-1994}.
\end{rems}

%In the sequel, we will also deal with Dirichlet heat kernels $k_D$ on finite
%unions $D = \bigcup_{j=1}^k D_j$ of regular domains with disjoint closures
%$\overline{D_j}$. We refer to those sets $D$ as {\em regular sets in $X$}.
%$k_D$ is then given by
%\[ k_D(t,x,y) = \begin{cases} k_{D_j}(t,x,y), & \text{if} \ x, y \in D_j, \\
%0, & \text{if $x, y$ are in different components}. \end{cases} \]

The following lemma states, in concise form, the crucial fact about heat 
kernels, which is needed in the next section.

\begin{lem}[Heat Kernel Lemma] \label{hkhelp}
  Let $\{ A_n \}$ and $\{ D_n \}$ be two sequences of subsets of $X$
  satisfying both the isoperimetric property {\rm {\bf (P)}}. Moreover, we assume
  that the sets $D_n$ are regular and that either $\{ A_n \}$ is an
  approximation of $\{ D_n \}$, or vice versa. Moreover, let $H^\omega
  = \Delta + V^\omega$, $\omega \in \Omega$, be a family of operators
  satisfying {\rm (\ref{potreg1})}. Then we have, for $n \to \infty$,
  \begin{equation} \label{heatconv} \sup_{\omega \in \Omega} \left |
  \frac{1}{|A_n|} \int_{A_n} k^\omega (t,x,x) dx - \frac{1}{|D_n|}
  \int_{D_n} k^\omega_{D_n} (t,x,x) dx \right | \longrightarrow 0.
  \end{equation}
\end{lem}

\begin{rem} Note that (\ref{heatconv}) can be also interpreted as the following 
limit of traces:
\[ \sup_{\omega \in \Omega} \left | \frac{1}{|A_n|} {\rm Tr} (\chi_{A_n} 
e^{-t H^\omega}) - \frac{1}{|D_n|} {\rm Tr} (e^{-t H^\omega_{D_n}}) \right 
  | \longrightarrow 0,
\]   
where $\chi_A$ denotes the characteristic function of $A \subset X$.  
\end{rem}

\begin{proof} Proposition \ref{kernellem} b) and property {\bf (P)} easily imply that
\[ \left | \frac{1}{|A_n|} \int_{A_n} k^\omega (t,x,x) dx - \frac{1}{|D_n|} 
\int_{D_n} k^\omega (t,x,x) dx \right | \longrightarrow 0. \] 
Thus we have to prove (\ref{heatconv}) only in the case $A_n = D_n$. Using,
again, Proposition \ref{kernellem} and property {\bf (P)} we conclude, for $h =
h(t,\epsilon)$, that
\begin{multline*}
\left | \frac{1}{|D_n|} \int_{D_n} k^\omega (t,x,x) dx -
\frac{1}{|D_n|} \int_{D_n} k^\omega_{D_n}(t,x,x) dx \right | 
\\ 
\le
\frac{1}{|D_n|} \left( \int_{(D_n \backslash \partial_h D_n)} +
\int_{(D_n \cap \partial_h D_n)} \right) (k^\omega(t,x,x) -
k^\omega_{D_n}(t,x,x)) dx 
\\ 
\le 
\frac{| D_n \backslash 
\partial_h D_n |}{| D_n |} \, \epsilon +  \frac{| \partial_h D_n |}{| D_n |} C(t) 
\longrightarrow \epsilon.
\end{multline*}
This finishes the proof, since $\epsilon > 0$ was arbitrary. \end{proof}

\section{Proof of the main theorem}

In this section we present the proof of Theorem \ref{main}. We assume
that $\{ D_n \}$ is an admissible sequence of $X$, and that $I_n$ and
$A_n =\phi(I_n)$ are the associated sequences (see Definition \ref{admseq}).
Let $\{ H^\omega \}_{\omega \in \Omega}$ be an ergodic random family of
Schr\"odinger operators satisfying the regularity condition (\ref{potreg1}).
In order to show almost-sure-convergence of the normalized eigenvalue
counting functions $N^\omega_{D_n}$ to a non-random distribution function 
$N$ at all continuity points it suffices to prove pointwise convergence of the
corresponding Laplace-transformations. This fact is a consequence of the 
following lemma. Recall that a distribution function is a non-negative, 
left-continuous, monotone increasing function.

\begin{lem}[Pastur/\v Subin]
  Let $N_n $ be a sequence of distribution functions
  such that 
  \begin{enumerate} 
  \item[a)] there exists a $c \in \RR$ such that $N_n (\lambda ) =0$ for all 
  $\lambda \le c$ and $n \in \NN$,
  \item[b)] there exists a $C_1: \RR^+ \to \RR$ such that $\tilde N_n (t)
  := \int e^{-\lambda t} dN_n (\lambda) \le C_1(t)$ for all $n \in
  \NN$, $t > 0 $, 
  \item[c)] $\lim_{n \to \infty} \tilde N_n(t) =: \psi(t)$
  exists for all $t > 0$.  
  \end{enumerate} 
  Then the limit 
  \[ N(\lambda) := \lim_{n \to \infty} N_n (\lambda) \]
  exists at all continuity points. $N$ is, again, a distribution function, and 
  its Laplace transform is $\psi$.
\end{lem}

{\bfseries Proof of Theorem \ref{main}:}
Now let $t > 0$ be fixed. The Laplace-transforms of the normalized 
eigenvalue counting functions can be written in terms of heat kernels:
\begin{eqnarray}
\tilde N^\omega_{D_n}(t) &=& \int_{D_n} e^{-t \lambda}\,
dN^\omega_{D_n}(\lambda)\ =\ \frac{1}{| D_n |} {\rm Tr}(e^{-t 
H_{D_n}^\omega}) \nonumber \\
&=& \frac{1}{| D_n |} \int_{D_n} k_{D_n}^\omega(t,x,x) dx. \label{ggg} 
\end{eqnarray} 
Applying the Heat Kernel Lemma \ref{hkhelp}, we obtain
\[ \lim_{n \to \infty} \left | \tilde N^\omega_{D_n}(t) - \frac{1}{| A_n |} 
\int_{A_n} k^\omega(t,x,x) dx \right | = 0. \]
Note that the function $(\omega,x) \mapsto k^\omega(t,x,x)$ is jointly 
measurable. (\ref{potcomp}) and the spectral theorem imply that 
\[ e^{-t H^{T_\gamma \omega}} = U_\gamma e^{-t H^\omega} U_\gamma^*, \]
where $U_\gamma: L^2(X) \to L^2(X)$ are unitary operators, defined by
$U_\gamma f(x) = f(\gamma^{-1}x)$. This yields 
$k^{T_\gamma \omega}(t,x,x) = k^\omega(t,\gamma^{-1} x,\gamma^{-1} x)$, and
we can apply Lemma \ref{Lindenstrausslem}. Consequently, we have, 
for almost all $\omega \in \Omega$:  
\[
\lim_{n \to \infty} \tilde N^\omega_{D_n}(t) = \lim_{n \to \infty}   
\frac{1}{| A_n |} \int_{A_n} k^\omega(t,x,x) dx = \frac{1}{| \FF |} 
\EE \left( \int_\FF k^\bullet(t,x,x) dx \right).
\]
Note that the conditions of the Pastur-\v Subin Lemma are satisfied: The previous
considerations imply c), for almost all $\omega \in \Omega$. a) holds with
$c = - C_0$, where $C_0$ is the constant in (\ref{potreg1}), and b) 
follows from (\ref{ggg}) and Proposition \ref{kernellem}. Consequently, an
application of \v Subin's lemma finishes the proof of Theorem \ref{main}.
Moreover, we obtain an explicit formula for the Laplace transform of the 
non-random IDS:
\[ \tilde N(t) = \int e^{-\lambda t} dN(\lambda) = \frac{1}{| \FF |} \EE 
\left( \int_\FF k^\bullet(t,x,x) dx \right). \]

\section{Discussion}

There are at least two natural extensions of our results including
random higher order terms. From the physical point of view it would be
interesting to include a magnetic field term in the Schr\"odinger
operator. The non-Euclidean setting raises the question whether one
can consider the pure Laplace operator on a differentiable manifold
equipped with a family of Riemannian metrics depending ergodically on
a random parameter. This may model, e.g., a quantum mechanical system
of a membrane with random hollows. For these cases, as well as for
more singular potentials, we would need to extend the methods of
Section 3.

We restricted ourselves to the case where the group acton is {\em discrete}
and the configuration space is {\em continuous}. One could also consider 
actions of Lie groups on manifolds; or a graph instead of a manifold as the 
configuration space.

In collaboration with Daniel Lenz we currently investigate whether our
IDS coincides with a trace of an appropriate von Neumann algebra. The
concept of a von Neumann algebra of random operators may be also
useful as a common abstract setting for all the situations described in
the previous paragraph. Such an abstract setting for the case of an abelian
group acting on another abelian group was studied, e.g., in \cite{Lenz}.

Furthermore, from the physical point of view, it would be interesting to
investigate finer properties of the IDS for particular models: the continuity 
or differentiability of $\lambda \mapsto N(\lambda)$, and the asymptotic 
behaviour as $\lambda$ approaches an edge of the spectrum of $\{ H^\omega \}$, 
cf.~\cite{Sz-89,Sz-90}.


\begin{thebibliography}{10}

\bibitem[Ad-93]{Adachi-93}
T. Adachi.
\newblock {\em A note on the {F{\o}lner} condition for amenability}
\newblock (dedicated to Professor Nobunori Ikebe).
\newblock Nagoya Math. J., 131:67--74, 1993.

\bibitem[AS-93]{AdachiS-93}
T. Adachi and T. Sunada.
\newblock {\em Density of states in spectral geometry}.
\newblock Comment. Math. Helv., 68:480--493, 1993.

\bibitem[Ba-72]{Ba}
H. Bass.
\newblock {\em The degree of polynomial growth of finitely generated nilpotent
groups}.
\newblock Proc. London Math. Soc., 25:603--614, 1972.

\bibitem[BB-93]{BAssB-1993}
R.~F. Bass, K. Burdzy. 
\newblock {\em On domain monotonicity of the Neumann heat kernel}.
\newblock J. Funct. Anal., 116:215--224, 1993.

\bibitem[BBEE+84]{Bonch-BruevichEEKMZ-1984}
V.~L. Bonch-Bruevich, R. Enderlein, B. Esser, R. Keiper, A.~G. Mirnov, 
and I.~P. Zyvagin.
\newblock {\em Elektronentheorie ungeordneter Halbleiter}.
\newblock VEB Deutscher Verlag der Wissenschaften, 1984.
\newblock Russian original: Moskau, Nauka, 1981.

\bibitem[BS-92]{BS}
J. Br\"uning and T. Sunada.
\newblock {\em On the spectrum of periodic elliptic operators}.
\newblock Nagoya Math. J., 126:159--171, 1992.

\bibitem[CL-90]{CarmonaL-90}
R. Carmona and J. Lacroix.
\newblock {\em Spectral Theory of Random {Schr\"odinger} Operators}.
\newblock Birkh\"auser, Boston, 1990.

\bibitem[CZ-94]{CarmonaZ-1994}
R. Carmona and W. Zheng.
\newblock {\em Reflecting {Brownian} motions and comparison theorems for 
{Neumann} heat kernels}.
\newblock J. Funct. Anal., 123(1):109--128, 1994.

\bibitem[Cha-84]{Cha}
I. Chavel.
\newblock {\em Eigenvalues in Riemannian geometry}.
\newblock Academic Press, New York, 1984.

\bibitem[Cha-86]{Chavel-1986}
I. Chavel. 
\newblock {\em Heat diffusion in insulated convex domains}. 
\newblock J. London Math. Soc., 34:473--478, 1986.

\bibitem[CGT-82]{CHGrT-1982}
J. Cheeger, M. Gromov and M. Taylor.
\newblock {\em Finite propagation speed, kernel estimates for functions of the
Laplace operator, and the geometry of complete Riemannian manifolds}.
\newblock J. Diff. Geom., 17:15--53, 1982.

\bibitem[Dav90]{Dav-90}
E.~B. Davies.
\newblock {\em Heat kernels and spectral theory}.
\newblock Cambridge University Press, Cambridge, 1990.

\bibitem[DM-97]{DodzM-97}
J. Dodziuk and V. Mathai.
\newblock {\em Approximating $L^2$ invariants of amenable covering spaces: 
A heat kernel approach}.
\newblock in ``Lipa's Legacy'', Contemp. Math., 211:151--167, 1997.

\bibitem[ES-84]{EfrosS-84}
A.~L. Efros and B.~I. Shklovski.
\newblock {\em Electronic Properties of Doped Semi-conductors}.
\newblock Springer, Berlin, 1984.

\bibitem[Elw-82]{Elworthy}
K.~D. Elworthy.
\newblock {\em Stochastic differential equations on manifolds}.
\newblock London Math. Soc. Lecture Note Series 70. 
\newblock Cambridge University Press, 1982.

\bibitem[Gro-81]{Gr}
M. Gromov.
\newblock {\em Groups of polynomial growth and expanding maps}.
\newblock IHES Publ. Math., 53:53--73, 1981.

\bibitem[KP-99]{KP}
L. Karp and N. Peyerimhoff.
\newblock {\em Horospherical means and uniform distribution of curves}.
\newblock Math. Z., 231:655--677, 1999.

\bibitem[Kir-89]{Kirsch-89a}
W. Kirsch.
\newblock {\em Random {Schr\"odinger} operators}.
\newblock In H.~Holden and A.~Jensen, editors, {\em {Schr\"odinger} Operators},
  Lecture Notes in Physics, 345, Springer, Berlin, 1989.

\bibitem[KM-82a]{KirschM-82a}
W. Kirsch and F. Martinelli.
\newblock {\em On the ergodic properties of the spectrum of general random
  operators}.
\newblock J. Reine Angew. Math., 334:141--156, 1982.

\bibitem[KM-82b]{KiMa-82}
W. Kirsch and F. Martinelli.
\newblock {\em On the density of states of Schr\"odinger operators with a 
random potential}.
\newblock J. Phys. A: Math. Gen., 15:2139--2156, 1982.

\bibitem[Kre-85]{Krengel-85}
U. Krengel.
\newblock {\em Ergodic Theorems}.
\newblock With a Supplement by Antoine Brunel.
\newblock Studies in Mathematics. Walter de Gruyter, 1985.

\bibitem[Le-99]{Lenz}
D. Lenz.
\newblock {\em Random operators and crossed products}.
\newblock Mathematical Physics, Analysis and Geometry, 2:197--220, 1999.

\bibitem[LY-86]{LiY-86}
P. Li and S.~T. Yau.
\newblock {\em On the parabolic kernel of the {Schr\"odinger} operator}.
\newblock Acta Mathematica, 156:153--201, 1986.

\bibitem[LGP-88]{LifshitzGP-88}
I.~M. Lifshitz, S.~A. Gredeskul, and L.~A. Pastur.
\newblock {\em Introduction to the Theory of Disordered Systems}.
\newblock Wiley, New York, 1988.
\newblock Russian original: Nauka, Moscow, 1982.

\bibitem[Lif-85]{Lifschitz-1985}
Lifschitz {Memorial} {Issue}.
\newblock J. Statist. Phys., 38(1--2), 1985.

\bibitem[Lin-99]{Lindenstrauss-99}
E. Lindenstrauss.
\newblock {\em Pointwise theorems for amenable groups}.
\newblock Electr. research announcem. of the Amer. Math. Soc., 
5:82--90, 1999.

\bibitem[LS-99]{LueckS-1999}
W. L\"uck and T. Schick.
\newblock {\em $L^2$-{Torsion} of hyperbolic manifolds of finite volume}.
\newblock Geom. Funct. Anal., 9:518--567, 1999. 

\bibitem[Mil-68]{Milnor}
J. Milnor.
\newblock {\em A note on curvature and fundamental group}.
\newblock J. Differ. Geom., 2:1--7, 1968. 

\bibitem[Pas-71a]{Pas-71a}
L.~A. Pastur.
\newblock {\em The Schr\"odinger operator with random potential}.
\newblock Theoret. and Math. Phys., 6:415--424, 1971. 

\bibitem[Pas-71b]{Pas-71b}
L.~A. Pastur.
\newblock {\em Selfaverageability of the number of states of the Schr\"odinger
equation with a random potential}.
\newblock Mat. Fiz. i Funkcional. Anal., 238:111--116, 1971.

\bibitem[Pas-80]{Pas-80} L.~A. Pastur.  
\newblock {\em Spectral properties of disordered systems in the one-body 
approximation}. 
\newblock Commun. Math. Phys., 75:17-9-196, 1980.

\bibitem[PF-92]{PasturF-92}
L.~A. Pastur and A.~L. Figotin.
\newblock {\em Spectra of Random and Almost-Periodic Operators}.
\newblock Springer Verlag, Berlin, 1992.

\bibitem[PV-00]{PV-00}
N. Peyerimhoff and I. Veseli{\'c}.
\newblock {\em Integrated density of states for random Schr\"odinger 
operators on manifolds}.
\newblock MaPhySto Research Report, No. 33, 2000.

\bibitem[RSII-75]{ReedS-II}
M. Reed and B. Simon.
\newblock {\em Methods of modern mathematical physics}.
\newblock Vol. II: Fourier analysis, self-adjointness.
\newblock Academic Press, Boston, 1975.

\bibitem[Shu-79]{Shub-79}
M.~A. \v Subin.
\newblock {\em Spectral theory and the index of elliptic operators with 
almost-periodic coefficients}. 
\newblock Russian Math. Surveys, 34:109--158, 1979.

\bibitem[Shu-82]{Shub-82}
M.~A. \v Subin.
\newblock {\em Density of states of self adjoint operators with almost periodic
coefficients}.
\newblock Amer. Math. Soc. Translations, 118:307--339, 1982. 

\bibitem[Shul-88]{Shul-88}
A. Shulman.
\newblock {\em Maximal ergodic theorems on groups}.
\newblock Dep. Lit. NIINTI, No. 2184, 1988.

\bibitem[Sim-71]{Simon-1971}
B. Simon.
\newblock {\em Quantum mechanics for {H}amiltonians defined as quadratic
  forms}.
\newblock Princeton Series in Physics.
\newblock Princeton University Press, Princeton, N. J., 1971.

\bibitem[Sz-89]{Sz-89}
A.-S. Sznitman.
\newblock {\em Lifschitz tail and Wiener sausage on hyperbolic space}.
\newblock Comm. Pure Appl. Math., 42:1033--1065, 1989.

\bibitem[Sz-90]{Sz-90}
A.-S. Sznitman.
\newblock {\em Lifschitz tail on hyperbolic space: Neumann conditions}.
\newblock Comm. Pure Appl. Math., 43:1--30, 1990.

\bibitem[Tay-96]{Taylor-1996}
M.~E. Taylor.
\newblock {\em Partial differential equations {I}: Basic theory}.
\newblock Springer-Verlag, New York, 1996.

\bibitem[Tem-72]{Tempelman-72}
A. Tempelman.
\newblock {\em Ergodic theorems for general dynamical sytems}.
\newblock Trans. Moscow Math. Soc., 26:95--132, 1972.


\bibitem[Tem-92]{Tempelman-92}
A. Tempelman.
\newblock {\em Ergodic theorems for group actions. Informational and
  thermodynamical aspects}, volume 78 of {\em Mathematics and its
  Applications}.
\newblock Kluwer Academic Publishers, 1992.

\bibitem[Thu-97]{Th}
W.~P. Thurston.
\newblock {\em Three-dimensional geometry and topology}.
\newblock Princeton University Press, 1997. 

\end{thebibliography}
\end {document}